\documentclass{Interspeech}

\interspeechcameraready

\usepackage{microtype}
\usepackage{xcolor}

\usepackage{booktabs}
\usepackage{multirow}
\usepackage{tabularx}
\newcommand{\mytable}{
    \eightpt
    \centering
    \renewcommand{\arraystretch}{1.2}
}
\newcolumntype{C}{>{\centering\arraybackslash}X}
\newcolumntype{L}{>{\raggedright\arraybackslash}X}
\newcolumntype{R}{>{\raggedleft\arraybackslash}X}
\newcolumntype{P}[1]{>{\raggedright\arraybackslash}p{#1}}
\usepackage{siunitx}
\usepackage{etoolbox}                           %
\newcommand{\ubold}{\fontseries{b}\selectfont}  %
\robustify\ubold                                %
\newcommand{\tablecaptionsep}{\vspace{-5pt}}
\newcommand{\ii}[1]{{\scriptsize \textcolor{gray}{#1}}} 

\graphicspath{{fig/}}

\DeclareMathOperator*{\argmin}{arg\,min}

\usepackage{cite}
\title{LinearVC: Linear transformations of self-supervised features \\ through the lens of voice conversion}
\author[affiliation={1}]{Herman}{Kamper}
\author[affiliation={1}]{Benjamin}{van Niekerk}
\author[affiliation={2}]{Julian}{Zaïdi}
\author[affiliation={2}]{Marc-André}{Carbonneau}
\affiliation{Electrical and Electronic Engineering}{Stellenbosch University}{South Africa}
\affiliation{Ubisoft La Forge}{Montreal}{Canada}
\email{kamperh@sun.ac.za}

\begin{document}
    
    \maketitle
    
    \begin{abstract}
        We introduce LinearVC, a simple voice conversion method that sheds light on the structure of self-supervised representations. First, we show that simple linear transformations of self-supervised features effectively convert voices. Next, we probe the geometry of the feature space by constraining the set of allowed transformations. We find that just rotating the features is sufficient for high-quality voice conversion. This suggests that content information is embedded in a low-dimensional subspace which can be linearly transformed to produce a target voice. To validate this hypothesis, we finally propose a method that explicitly factorizes content and speaker information using singular value decomposition; the resulting linear projection with a rank of just 100 gives competitive conversion results. Our work has implications for both practical voice conversion and a broader understanding of self-supervised speech representations. Samples and code: {\url{https://www.kamperh.com/linearvc/}}.
    \end{abstract}

    \section{Introduction}

    Voice conversion aims to alter input speech to mimic a target speaker's voice~\cite{mohammadi2017overview,sisman2020overview}. 
    It has applications in entertainment~\cite{disney2021mandolarian}, language tutoring~\cite{geng2024pilot}, accessible speech processing~\cite{singer2021respeecher,baas2023bvoice,hajal2025unsupervised}, and  anonymization~\cite{panariello2024voiceprivacy}.
    This wide range of applications has naturally led to diverse solutions.
    Some systems leverage large spoken language models, prompting them to generate speech in a desired voice \cite{han2024vall, soundstrom}, while others rely on speaker embeddings as a conditioning signal~\cite{guo2023quickvc,yang2024streamvc}. 
    Recent studies have achieved competitive results by applying simple methods, such as nearest neighbours, on top of self-supervised learned (SSL) representations~\cite{lin2021fragmentvc,baas2023voice,baas2023bvoice,asadulaev2024}.
    The success of these simple voice conversion systems stems directly from the internal structure of the SSL features used.
    But we still only have a cursory understanding of how SSL models organize content and speaker information. %

    In this paper we propose LinearVC, a simple method that also gives new insights into the structure of SSL features.
    The method uses features from an intermediate layer of WavLM~\cite{chen2022wavlm}, an established SSL system.
    Encoded source and target frames are paired using nearest neighbours to form a training set.
    A linear projection is then learned between source and target frames.
    At inference, source speech is linearly projected and then vocoded using a pretrained vocoder, yielding the converted waveform.
    This approach is heavily inspired by k-nearest neighbours voice conversion (kNN-VC)~\cite{baas2023voice}. 
    Here we replace kNN-VC's non-linear nearest neighbour mapping with a linear transformation.

    We start by showing that LinearVC gives comparable performance to 
    state-of-the-art systems such as
    kNN-VC
    and SoundStorm~\cite{soundstrom}.
    This finding---that a linear mapping can convert voices---suggests that SSL features implicitly disentangle phonetic and speaker identity information. %
    This aligns with~\cite{liu2023self,mohamed2024orthogonality},
    which showed that phonetic and speaker information is encoded in orthogonal directions in SSL representations.
    To further investigate how the SSL space is organized, we %
    constrain the types of linear transformations allowed.
    With just rotation and reflection, high-quality voice conversion is possible, suggesting 
    that phonetic information resides in a lower-dimensional subspace that is maintained from one speaker to another.
    
    To show this conclusively, we finally propose a method that explicitly factorizes content and speaker information.
    We use singular value decomposition to find a content representation that is shared across speakers, together with a set of speaker-specific linear transformations.
    For voice conversion, new input is projected into the common content space and then projected out to the desired speaker---a version of LinearVC with a low-rank linear projection. We find that phonetic content can be encoded with a rank as low as 16, but voice conversion is poor. When we increase the rank to around 100, content and speaker information are retained, yielding competitive conversion results.
    
    Many studies have examined how information is encoded across the different layers of SSL models~\cite{pasad2021layer,mousavi2024audiotokens}.
    Our work shows 
    how SSL models also encode different types of information in different subspaces within the same layer.
    Additionally, it complements work on orthogonality in SSL representations~\cite{liu2023self,mohamed2024orthogonality}, offering deeper insights into the geometry of the SSL space.
    Finally, our findings help explain the success of SSL-based voice conversion methods such as \cite{lin2021fragmentvc,
        baas2023bvoice,vanniekerk_softvc,baas2023voice,shan2024phoneme,asadulaev2024}.
    LinearVC is therefore a straightforward method that not only achieves strong voice conversion results but also provides new insights into the structure and use of SSL representations.
    
    \begin{figure*}[!t]
        \centering
        \includegraphics[scale=0.66666]{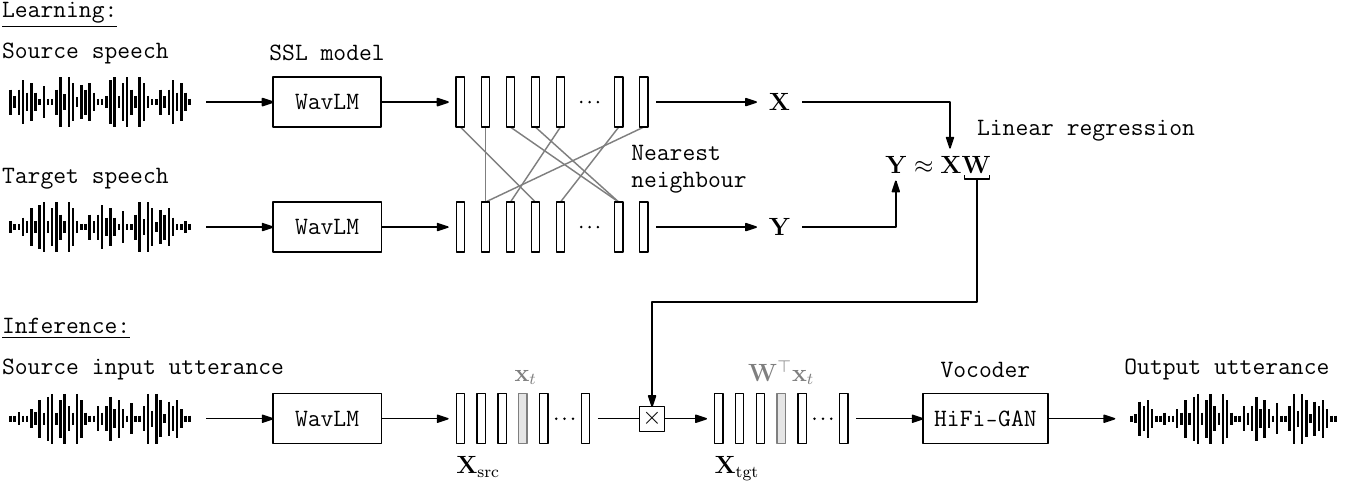}
        \caption{
            LinearVC: At learning time (top), source speech frames are paired with target frames by finding their single nearest neighbours. %
            A projection matrix is learned by finding the least squares solution for mapping from source to target frames. At inference time (bottom), frames from an unseen source utterance are linearly projected and then vocoded to get speech in the target speaker's voice.
        }
        \label{fig:method}
    \end{figure*}

    \section{LinearVC}
    \label{sec:linearvc}

    The LinearVC framework is illustrated in Figure~\ref{fig:method}.
    During training (top), a linear transformation from a source to a target speaker is learned.
    Utterances are first encoded into $D$-dimensional feature frames using a large SSL speech model, like WavLM~\cite{chen2022wavlm}.
    Then, for each of the $N$ source frames, we find the closest neighbour from the set of $M$ target frames.
    The source frames are arranged in a matrix $\mathbf{X} \in \mathbb{R}^{N \times D}$ and the corresponding target frames in a matrix $\mathbf{Y} \in \mathbb{R}^{N \times D}$.
    We then find a projection matrix $\mathbf{W}$ by solving the multivariate linear regression problem:
    \begin{equation}
        \argmin_{\mathbf{W}} \lVert \mathbf{Y} - \mathbf{X} \mathbf{W} \rVert_{\textrm{F}}^2
        \label{eq:lr}
    \end{equation}
    where $\lVert \cdot \rVert_{\textrm{F}}$ is the Frobenius norm.
    At inference time (Figure~\ref{fig:method}-bottom), each frame in a source utterance $\mathbf{X}_{\textrm{src}}$ is linearly projected to get the converted output ${\mathbf{X}}_{\textrm{tgt}} = \mathbf{X}_{\textrm{src}} \mathbf{W}$.
    A pre-trained vocoder (we use HiFiGAN~\cite{kong2020hifi}) produces the final speech waveform from the projected frames.
    
    The intuition behind this approach is that phonetic information is structured in the SSL space in a way that is common to all speakers; if such a phonetic subspace exists, then projecting these into different locations in the space should alter voice characteristics while maintaining content.
    Support for our intuition comes from \cite{cho2024self}, for instance, which showed that acoustic-to-articulation models can be transferred between speakers using a linear transformation, and~\cite{asadulaev2024}, which improved an optimal transport-based voice conversion approach using linear interpolations between latent source and target representations.
    
    In the next section, we confirm our hypothesis that learned linear transformations of SSL features can convert voices.

    \section{Voice conversion experiments}
    \label{sec:experiments}
    
    We compare LinearVC to three state-of-the-art voice conversion methods, demonstrating its effectiveness despite its simplicity.
    
    \subsection{Experimental setup}
    \label{sec:experimental_setup}
    
    \hspace{\parindent}\textbf{Data.}
    Experiments are performed on the English Libri\-Speech corpus~\cite{panayotov2015librispeech}. %
    We perform development experiments on the dev-clean subset and report final scores on the test-clean subset.
    Each subset has speech from 40 speakers.
    For learning LinearVC's projection matrix and for conditioning kNN-VC, we use 2.7~minutes of audio from each speaker.
    For evaluations, we sample five utterances from each of the speakers.
    The evaluation utterances do not overlap with the reference data.

    \textbf{LinearVC implementation.}
    LinearVC operates on features from the sixth layer of WavLM-Large~\cite{chen2022wavlm}, giving frames with a dimensionality of $D = \text{1024}$ every 20~ms for 16~kHz %
    input. 
    We chose WavLM based on its performance in~\cite{pasad2023comparative,mousavi2024dasb} %
    and use layer six
    based on~\cite{baas2023voice}. %
    In development experiments, we considered different configurations for LinearVC's matching step (Figure~\ref{fig:method}-top) and found that a single nearest neighbour with cosine distance worked best.
    For vocoding, we use a universal HiFi-GAN~\cite{kong2020hifi} operating on WavLM features; it is trained on LibriSpeech train-clean-100 using the hyperparameters from~\cite{kong2020hifi} with the data augmentation strategy from~\cite{baas2023voice}. %
    
    \textbf{Systems we compare to.}
    We compare LinearVC to three existing voice conversion systems.
    The first is kNN-VC~\cite{baas2023voice}. %
    Similar to LinearVC's learning phase, kNN-VC
    maps each frame in a source utterance to its nearest neighbour in target speech data; the mapped frames are then directly fed to a neural vocoder.
    Second is FreeVC, which uses a variational autoencoder with data augmentation to discard speaker information~\cite{li2023freevc}.
    Third, we use SoundStorm~\cite{soundstrom}, a representative large spoken language model. 
    It is a non-autoregressive transformer 
    which can be conditioned using a prompt of a few seconds from the target speaker.
    We use our own SoundStorm implementation, trained on 60k hours from LibriLight~\cite{librilight}, excluding the evaluation~data.

    \textbf{Objective evaluations.}
    Each evaluation utterance is converted to all speakers other than the source, giving an evaluation based on 7800 conversions per model.
    As in ~\cite{yi2020voice}, we measure intelligibility by computing the word/character error rate~(W/CER) between the output of a speech recognition system (Whisper-small~\cite{radford2023robust}) applied to the converted speech and the ground truth transcriptions.
    Lower W/CER indicates better intelligibility.
    To measure speaker similarity, we use the approach from~\cite{das2020predictions}, where a speaker verification system~\cite{snyder2018x} must discriminate between real and converted speech from a target speaker.
    A higher equal error rate~(EER) corresponds to a better VC system, with 50\% indicating conversions that fool the verifier perfectly.

    \textbf{Subjective evaluations.} 
    We evaluate naturalness and speaker similarity using  MUSHRA-like listening tests~\cite{ITU2015}.
    Evaluation utterances of roughly five seconds are converted to ten target speakers, half matching the gender of the source and half being cross-gender.
    Each conversion is evaluated by at least 15 raters, after excluding cases where low-quality anchors are rated above other stimuli.
    To evaluate speaker similarity, raters compare converted utterances to a real reference utterance from the target speaker; the source utterance is used as a low anchor.
    For naturalness, there is no reference, and a degraded synthesized utterance gives a low anchor.
    We report mean scores with 95\% confidence intervals.
    Following \cite{StatMUSHRA}, we use the Friedman test~\cite{Friedman1937} to determine whether at least one system differs significantly from the others. For cases that differ, we perform pairwise comparisons using Wilcoxon signed-rank tests~\cite{Wilcoxon1945} with Bonferroni correction.

    \begin{table}[!t]
        \caption{Intelligibility (W/CER), naturalness, and speaker similarity (EER, similarity) results (\%) on LibriSpeech test-clean.}
        \tablecaptionsep
        \label{tbl:results_sota}
        \mytable
        \setlength{\tabcolsep}{4pt}  %
        \begin{tabularx}{\linewidth}{
                @{}
                L
                c
                c
                c
                cc
                @{}
            }
            \toprule
            Model & {WER$\downarrow$} & {CER$\downarrow$} & {EER$\uparrow$} & {Natural$\uparrow$} & {Similarity$\uparrow$} \\
            \midrule
            Ground truth & 4.3 & 2.3 & {-} & {-} & {-} \\
            \addlinespace
kNN-VC~\cite{baas2023voice} & 5.7 & 2.9 & \ubold 38.9 & 60.6$\pm$3.6 & 67.2$\pm$2.7\\
FreeVC~\cite{li2023freevc} & 5.7 & 3.0 & 10.5 & \textbf{71.1$\pm$3.6} & 48.7$\pm$2.9\\
SoundStorm~\cite{soundstrom} & \ubold 4.6 & \ubold 2.4 & 30.2 & 58.6$\pm$4.0 & \textbf{68.6$\pm$3.2} \\
LinearVC  & 4.9 & 2.6 & 33.6 & 62.5$\pm$3.5 & 67.5$\pm$2.6\\
LinearVC factor.\  & 4.7 & 2.5 & 35.2 & 62.3$\pm$3.7 & 64.2$\pm$3.1\\            
            \bottomrule
        \end{tabularx}
    \end{table}
    
    \begin{figure*}[!t]
        \centering
        \includegraphics[scale=0.66666]{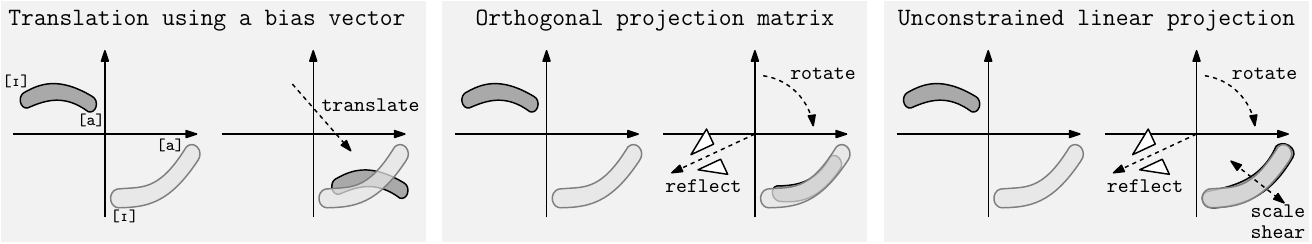}
        \caption{A cartoon illustration of the effects of different linear transformations on the representation space. The dark and light blobs respectively represent where features from two distinct speakers are situated in the SSL space.}
        \label{fig:transforms}
    \end{figure*}
    
    \subsection{Voice conversion results}
    
    Table~\ref{tbl:results_sota} presents objective and subjective voice conversion results on LibriSpeech test-clean.
    All systems produce intelligible output, with low W/CERs approaching that of ground truth speech.
    LinearVC performs comparably to SoundStorm and kNN-VC, with no statistically significant differences in naturalness or speaker similarity (confirmed through significance tests). In contrast, FreeVC achieves significantly higher naturalness, but at the cost of a significantly lower speaker similarity, as reflected in both the subjective evaluation and the EER of 10.5\%.
    Interestingly, EER only gives a coarse correspondence with perceived speaker similarity, so small differences in EER should be taken with a grain of salt.

    In short, using a single linear projection, the proposed LinearVC system achieves voice conversion with a good compromise between intelligibility, naturalness, and speaker similarity. This indicates that SSL features implicitly disentangle phonetic information from speaker identity. We further validate this in the next sections, and then introduce a factorized version of LinearVC that explicitly disentangles content from speaker identity through low-rank factorization. (Results for this factorized system are already given in the last row of Table~\ref{tbl:results_sota}.)

    \section{Further analysis: Constrained linear transformations}
    \label{sec:constraints}

    We now constrain the type of transformation in LinearVC to gain insights into the organization of the SSL representation space.
    Figure~\ref{fig:transforms} gives a cartoon illustration of the different configurations.
    First, we exclusively use a bias vector, allowing only for translation.
    Then, $\mathbf{W}$ is constrained to be orthogonal, allowing only for rotations and reflections.
    In this case, $\mathbf{W}$ in~\eqref{eq:lr} is found using~\cite{schonemann1966generalized} to solve the orthogonal Procrustes problem.
    Finally, no constraints are placed on $\mathbf{W}$: we can have translation, rotation, reflection, scaling, and shearing.
    This is what we have done so far, except that up to this point we did not include a bias (so no translations were used in the experiments in Section~\ref{sec:experiments}). 
    
    Table~\ref{tbl:linear_transforms} gives results on LibriSpeech dev-clean.
    Regardless of the transformation, intelligibility (W/CER) is maintained, i.e.\ phonetic content is preserved.
    This is evidence of a phonetic subspace structure shared across speakers, supporting the (sometimes implicit) hypothesis in other work~\cite{vanniekerk_softvc,liu2023self} that SSL models internally disentangle phonetic content from speaker identity.
    The results from a simple translation (row 2) are surprising: listening to samples reveals that the voice changes considerably while intelligibility is maintained (low W/CER).
    Although the voice changes, it does not mimic the target well (low EER).
    But the EER still gets close to the FreeVC baseline in  Table~\ref{tbl:results_sota}, simply through translation in the SSL space.

    \begin{table}[!t]
        \caption{Intelligibility (W/CER) and speaker similarity (EER) results (\%) for different linear transformations on dev-clean.}
        \label{tbl:linear_transforms}
        \tablecaptionsep
        \mytable
        \setlength{\tabcolsep}{4pt}  %
        \begin{tabularx}{\linewidth}{
                @{}
                l
                l
                c
                C
                S[table-format=2.1]
                @{}
            }
            \toprule
            & Model                           & {WER$\downarrow$} & {CER$\downarrow$} & {EER$\uparrow$} \\
            \midrule
            \ii{1} & Ground truth                    & 4.1             & 2.1             & {-}          \\
            \addlinespace
            \ii{2} & Bias only                       & 5.0 & 2.9 & 7.7           \\
            \ii{3} & Orthogonal                      & 5.1             & 2.9             & 27.7          \\
            \ii{4} & Orthogonal with bias            & 5.0 & 2.9 & 28.3 \\
            \ii{5} & No constraints     & 5.4             & 3.2             & 31.8          \\
            \ii{6} & No constraints with bias & 5.5 &  3.2 & 31.8   \\        
            \bottomrule
        \end{tabularx}
    \end{table}
    
    Allowing for rotation and reflection (rows 3 and 4) greatly improves speaker similarity (high EER).
    Adding scaling and shearing (rows 5 and 6) provide further boosts---but these are minimal.
    Adding a bias (rows 4 and 6) also does not greatly affect performance.
    It appears, then, that rotations and reflections around the origin are the main contributors to the competitive voice conversion performance of LinearVC.
    This again supports the hypothesis that content and speaker is captured in different subspaces---this is what allows us to alter the voice without changing content through linear manipulations. %
    
    \begin{figure}[!t]
        \centering
        \includegraphics[width=0.8\linewidth]{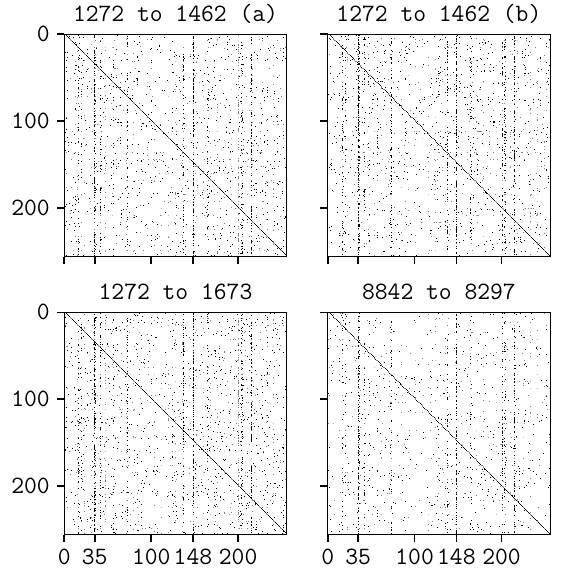}    
        \caption{
            Projection matrices for different speaker pairs.
            Top row: Same speaker pair, but using different speech subsets (a) and (b) for learning. %
            Matrices are thresholded and binarized, and only the 
            first 256 dimensions (out of 1024)
            are shown.
        }
        \label{fig:projmats}
    \end{figure}

    Before more formally showing that a common phonetic structure can be extracted, we qualitatively look at projection matrices to see if we can observe any commonalities.
    Figure~\ref{fig:projmats} shows unconstrained projection matrices, both for the same speaker pair but using different samples to estimate the matrices (top row) and between completely different speaker pairs (bottom row).
    To enhance matrix readability, a threshold was applied and absolute values binarized.
    Regardless of the exact samples used to learn $\mathbf{W}$ (top row), or the speaker pairs involved (bottom row), we see commonalities. E.g.\ the linear transformations consistently modify dimensions 35 and 148 using values from all the other dimensions.
    Since the result of the transformations is only a change in voice and not in content, we know that these modifications cause movement in the speaker subspace.

    \section{Factorizing out a  shared content subspace}

    The experiments in the previous section support the hypothesis that content information is embedded in a low-dimensional subspace, which can be linearly transformed to produce a target voice.
    In this section, we validate this interpretation by explicitly disentangling content and speaker information.
    Specifically, we factorize the SSL features into a content representation shared across all speakers and a set of speaker-specific transformations.
    The approach shares the spirit of very recent parallel work~\cite{ruggiero2025etawavlm}.
    
    \subsection{LinearVC with content factorization}
    
    First, we extract self-supervised features for $K$ distinct speakers.
    Then, we choose a single source speaker and find matching feature vectors from the other speakers using nearest neighbours.
    We arrange these features into matrices $\mathbf{X}_k \in \mathbb{R}^{N \times D}$, one for each speaker. 
    At this point, we have the same content spoken by different speakers.
    Next, we solve an optimization problem to factorize $\mathbf{X}_1, \ldots, \mathbf{X}_K$ into the product of a shared content representation $\mathbf{C}$ and a speaker-specific transformation $\mathbf{S}_k$:
    \begin{equation*}
        \begin{aligned}
            \underset{\mathbf{C}, \mathbf{S}_k}{\text{min}} &\quad \sum_{k = 1}^K \lVert \mathbf{X}_k - \mathbf{C} \mathbf{S}_k \rVert _F^2 \\
            \text{subject to } &\quad \text{rank}(\mathbf{C} \mathbf{S}_k) \leq r \, %
        \end{aligned}
    \end{equation*}
    where $r$ is a hyperparameter constraining the rank of the factorization.
    We can rewrite this problem %
    by concatenating the matrices $\mathbf{X}_k$ along their feature dimensions, giving:
    \[
    \underset{\mathbf{C}, \mathbf{S}}{\text{min}} \, \lVert \mathbf{X} - \mathbf{C} \mathbf{S} \rVert _F^2 \quad \text{subject to} \quad \text{rank}(\mathbf{C} \mathbf{S}) \leq r
    \]
    where $\mathbf{X} \in \mathbb{R}^{N \times KD}$ and $\mathbf{S} \in \mathbb{R}^{r \times KD}$ are the resulting block matrices.\footnote{This is only strictly true if there are at least $r$ non-zero eigenvalues.}
    We can solve this problem through a singular value decomposition of the block matrix $\mathbf{X}$.
    Concretely, we approximate each $\mathbf{X}_k$ as $\mathbf{U} \mathbf{\Sigma} \mathbf{S}_k$, where $\mathbf{U} \in \mathbb{R}^{N \times r}$ is an orthogonal matrix, $\mathbf{\Sigma} \in \mathbb{R}^{r \times r}$ is a diagonal matrix of the $r$ largest singular values, and $\mathbf{S}_k \in \mathbb{R}^{r \times D}$ is the corresponding block of right-singular vectors of $\mathbf{X}$.
    The product $\mathbf{U} \mathbf{\Sigma}$ represents the shared content $\mathbf{C}$, and each $\mathbf{S}_k$ is a speaker-specific linear transformation.
    
    To perform conversion given this factorization, we project a source utterance $\mathbf{X}_\textrm{src}$ to the content subspace, and then apply the target speaker transformation.
    Since $\mathbf{X}_\textrm{src} \approx \mathbf{C} \mathbf{S}_\textrm{src}$, we can multiply by the pseudoinverse of the source speaker transformation  $\mathbf{S}_\textrm{src} ^+$ to project to the content subspace: $\mathbf{X}_\textrm{src} \mathbf{S}_\textrm{src}^+ \approx \mathbf{C}$.
    Then we apply the target speaker transformation to convert to the desired speaker's voice:
    \begin{equation*}
        \mathbf{X}_\textrm{tgt} = \mathbf{X}_\textrm{src} \mathbf{S}_\textrm{src}^+ \mathbf{S}_\textrm{tgt}
        \label{eq:factorization-conversion}
    \end{equation*}
    The $\mathbf{S}_\textrm{src}^+ \mathbf{S}_\textrm{tgt}$ term functions like the projection matrix $\mathbf{W}$ of the non-factorized LinearVC. The difference here is that we can explicitly set the transformation's rank $r$.
    
    \begin{figure}[!t]
        \centering
        \includegraphics[width=0.9\linewidth]{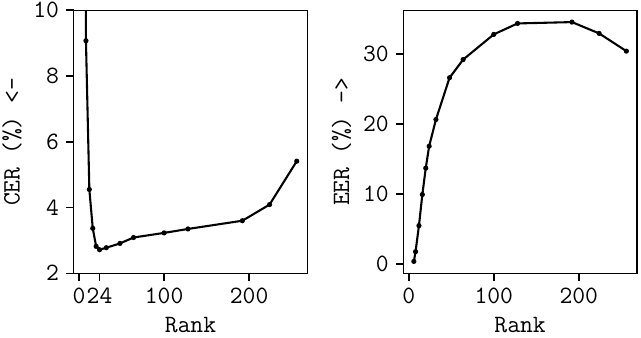}    
        \caption{Intelligibility (CER) and speaker similarity (EER) on dev-clean as a function of rank when using content factorization.} 
        \label{fig:content-factorization}
    \end{figure}
    
    \subsection{Voice conversion results}
    
    To evaluate this approach, we use the same experimental setup as in Section~\ref{sec:experimental_setup}, with the same data setup as for LinearVC.
    
    First, we investigate the effect of rank on the voice conversion quality.
    Figure~\ref{fig:content-factorization} plots intelligibility (CER) and speaker similarity (EER) as we vary the factorization's rank.
    The CER plot shows that content can be represented with a low rank: a CER of less than 4\% is achieved with just 16 dimensions, with the best intelligibility achieved at a rank of 24.
    This demonstrates that we can significantly compress the representation before degrading intelligibility, supporting our hypothesis that content information lies along a low-dimensional subspace.
    The right panel shows that EER (objective speaker similarity) increases with rank, suggesting that more parameters are required to model the speaker transformations.
    However, increasing the rank above 200 reduces EER, %
    likely because some source speaker information leaks into the content representation.
    
    Finally, we compare content factorization to the other systems in Table~\ref{tbl:results_sota}.
    Based on Figure~\ref{fig:content-factorization}, we set the rank to $r = 100$ as a compromise between intelligibility and speaker similarity. 
    Overall, our factorization method performs on par with the other approaches, with no statistically significant difference between the two variants of LinearVC.
    These results show that the factorization effectively disentangles speaker and content information. 
    
    \section{Conclusion}
    
    We introduced LinearVC, a simple yet effective method for voice conversion that uses a linear projection of SSL speech representations. 
    In addition to providing a practical solution to voice conversion without requiring complex model training, it offers valuable insights into the organization of SSL features.
    Our experiments revealed that phonetic and speaker identity information reside in distinct subspaces, and that simple linear transformations within these subspaces suffice for high-quality voice conversion. 
    Based on this, we proposed a new low-rank factorization approach for separating out common phonetic content from speaker identity information.
    Future work may extend our analyses to more SSL models (other than WavLM), and explore applications in other tasks such as speech anonymization and phonetic content extraction.

    \newpage
    \bibliography{ref}

\begin{thebibliography}{10}
\providecommand{\url}[1]{#1}
\csname url@samestyle\endcsname
\providecommand{\newblock}{\relax}
\providecommand{\bibinfo}[2]{#2}
\providecommand{\BIBentrySTDinterwordspacing}{\spaceskip=0pt\relax}
\providecommand{\BIBentryALTinterwordstretchfactor}{4}
\providecommand{\BIBentryALTinterwordspacing}{\spaceskip=\fontdimen2\font plus
\BIBentryALTinterwordstretchfactor\fontdimen3\font minus
  \fontdimen4\font\relax}
\providecommand{\BIBforeignlanguage}[2]{{%
\expandafter\ifx\csname l@#1\endcsname\relax
\typeout{** WARNING: IEEEtran.bst: No hyphenation pattern has been}%
\typeout{** loaded for the language `#1'. Using the pattern for}%
\typeout{** the default language instead.}%
\else
\language=\csname l@#1\endcsname
\fi
#2}}
\providecommand{\BIBdecl}{\relax}
\BIBdecl

\bibitem{mohammadi2017overview}
S.~H. Mohammadi and A.~Kain, ``An overview of voice conversion systems,''
  \emph{Speech Communication}, 2017.

\bibitem{sisman2020overview}
B.~Sisman, J.~Yamagishi, S.~King, and H.~Li, ``An overview of voice conversion
  and its challenges: From statistical modeling to deep learning,''
  \emph{IEEE/ACM Transactions on Audio, Speech, and Language Processing}, 2020.

\bibitem{disney2021mandolarian}
Disney, ``Making of {Season 2} finale,'' \emph{Disney Gallery: The
  Mandalorian}, 2021.

\bibitem{geng2024pilot}
H.~Geng, D.~Saito, and N.~Minematsu, ``A pilot study of applying
  sequence-to-sequence voice conversion to evaluate the intelligibility of {L2}
  speech using a native speaker's shadowings,'' in \emph{APSIPA ASC}, 2024.

\bibitem{singer2021respeecher}
G.~Singer, ``Respeecher gives voice to {Michael York} in healthcare
  initiative,'' \emph{Respeecher Blog}, 2021.

\bibitem{baas2023bvoice}
M.~Baas and H.~Kamper, ``Voice conversion for stuttered speech, instruments,
  unseen languages and textually described voices,'' \emph{Communications in
  Computer and Information Science}, 2023.

\bibitem{hajal2025unsupervised}
K.~E. Hajal, E.~Hermann, A.~Kulkarni, and M.~Magimai.-Doss, ``Unsupervised
  rhythm and voice conversion of dysarthric to healthy speech for {ASR},'' in
  \emph{SPADE}, 2025.

\bibitem{panariello2024voiceprivacy}
M.~Panariello, N.~Tomashenko, X.~Wang, X.~Miao, P.~Champion, H.~Nourtel,
  M.~Todisco, N.~Evans, E.~Vincent, and J.~Yamagishi, ``The {VoicePrivacy 2022
  Challenge}: Progress and perspectives in voice anonymisation,''
  \emph{IEEE/ACM Transactions on Audio, Speech, and Language Processing}, 2024.

\bibitem{han2024vall}
B.~Han, L.~Zhou, S.~Liu, S.~Chen, L.~Meng, Y.~Qian, Y.~Liu, S.~Zhao, J.~Li, and
  F.~Wei, ``{VALL-E R}: Robust and efficient zero-shot text-to-speech synthesis
  via monotonic alignment,'' \emph{arXiv preprint arXiv:2406.07855}, 2024.

\bibitem{soundstrom}
Z.~Borsos, M.~Sharifi, D.~Vincent, E.~Kharitonov, N.~Zeghidour, and
  M.~Tagliasacchi, ``{SoundStorm}: Efficient parallel audio generation,''
  \emph{arXiv preprint arXiv:2305.09636}, 2023.

\bibitem{guo2023quickvc}
H.~Guo, C.~Liu, C.~T. Ishi, and H.~Ishiguro, ``{QuickVC}: Any-to-many voice
  conversion using inverse short-time fourier transform for faster
  conversion,'' \emph{arXiv preprint arXiv:2302.08296}, 2023.

\bibitem{yang2024streamvc}
Y.~Yang, Y.~Kartynnik, Y.~Li, J.~Tang, X.~Li, G.~Sung, and M.~Grundmann,
  ``{StreamVC}: Real-time low-latency voice conversion,'' in \emph{ICASSP},
  2024.

\bibitem{lin2021fragmentvc}
Y.~Y. Lin, C.-M. Chien, J.-H. Lin, H.-y. Lee, and L.-s. Lee, ``{FragmentVC}:
  Any-to-any voice conversion by end-to-end extracting and fusing fine-grained
  voice fragments with attention,'' in \emph{ICASSP}, 2021.

\bibitem{baas2023voice}
M.~Baas, B.~{van Niekerk}, and H.~Kamper, ``Voice conversion with just nearest
  neighbors,'' in \emph{Interspeech}, 2023.

\bibitem{asadulaev2024}
A.~Asadulaev, R.~Korst, V.~Shutov, A.~Korotin, Y.~Grebnyak, V.~Egiazarian, and
  E.~Burnaev, ``Optimal transport maps are good voice converters,'' \emph{arXiv
  preprint arXiv:2411.02402}, 2024.

\bibitem{chen2022wavlm}
S.~Chen, C.~Wang, Z.~Chen, Y.~Wu, S.~Liu, Z.~Chen, J.~Li, N.~Kanda,
  T.~Yoshioka, X.~Xiao, J.~Wu, L.~Zhou, S.~Ren, Y.~Qian, Y.~Qian, J.~Wu,
  M.~Zeng, X.~Yu, , and F.~Wei, ``{WavLM}: Large-scale self-supervised
  pre-training for full stack speech processing,'' \emph{IEEE Journal of
  Selected Topics in Signal Processing}, 2022.

\bibitem{liu2023self}
O.~D. Liu, H.~Tang, and S.~Goldwater, ``Self-supervised predictive coding
  models encode speaker and phonetic information in orthogonal subspaces,'' in
  \emph{Interspeech}, 2023.

\bibitem{mohamed2024orthogonality}
M.~Mohamed, O.~D. Liu, H.~Tang, and S.~Goldwater, ``Orthogonality and isotropy
  of speaker and phonetic information in self-supervised speech
  representations,'' in \emph{Interspeech}, 2024.

\bibitem{pasad2021layer}
A.~Pasad, J.-C. Chou, and K.~Livescu, ``Layer-wise analysis of a
  self-supervised speech representation model,'' in \emph{ASRU}, 2021.

\bibitem{mousavi2024audiotokens}
P.~Mousavi, J.~Duret, S.~Zaiem, L.~D. Libera, A.~Ploujnikov, C.~Subakan, and
  M.~Ravanelli, ``How should we extract discrete audio tokens from
  self-supervised models?'' in \emph{Interspeech}, 2024.

\bibitem{vanniekerk_softvc}
B.~van Niekerk, M.-A. Carbonneau, J.~Zaïdi, M.~Baas, H.~Seuté, and H.~Kamper,
  ``A comparison of discrete and soft speech units for improved voice
  conversion,'' in \emph{ICASSP}, 2022.

\bibitem{shan2024phoneme}
S.~Shan, Y.~Li, A.~Banerjee, and J.~B. Oliva, ``Phoneme hallucinator: One-shot
  voice conversion via set expansion,'' in \emph{AAAI}, 2024.

\bibitem{kong2020hifi}
J.~Kong, J.~Kim, and J.~Bae, ``{HiFi-GAN}: Generative adversarial networks for
  efficient and high fidelity speech synthesis,'' in \emph{NeurIPS}, 2020.

\bibitem{cho2024self}
C.~J. Cho, A.~Mohamed, A.~W. Black, and G.~K. Anumanchipalli, ``Self-supervised
  models of speech infer universal articulatory kinematics,'' in \emph{ICASSP},
  2024.

\bibitem{panayotov2015librispeech}
V.~Panayotov, G.~Chen, D.~Povey, and S.~Khudanpur, ``{LibriSpeech}: An {ASR}
  corpus based on public domain audio books,'' in \emph{ICASSP}, 2015.

\bibitem{pasad2023comparative}
A.~Pasad, B.~Shi, and K.~Livescu, ``Comparative layer-wise analysis of
  self-supervised speech models,'' in \emph{ICASSP}, 2023.

\bibitem{mousavi2024dasb}
P.~Mousavi, L.~Della~Libera, J.~Duret, A.~Ploujnikov, C.~Subakan, and
  M.~Ravanelli, ``{DASB} -- {Discrete} audio and speech benchmark,''
  \emph{arXiv preprint arXiv:2406.14294}, 2024.

\bibitem{li2023freevc}
J.~Li, W.~Tu, and L.~Xiao, ``{FreeVC}: Towards high-quality text-free one-shot
  voice conversion,'' in \emph{ICASSP}, 2023.

\bibitem{librilight}
J.~{Kahn}, M.~{Rivière}, W.~{Zheng}, E.~{Kharitonov}, Q.~{Xu}, P.~E.
  {Mazaré}, J.~{Karadayi}, V.~{Liptchinsky}, R.~{Collobert}, C.~{Fuegen},
  T.~{Likhomanenko}, G.~{Synnaeve}, A.~{Joulin}, A.~{Mohamed}, and E.~{Dupoux},
  ``{Libri-Light}: A benchmark for {ASR} with limited or no supervision,'' in
  \emph{ICASSP}, 2020.

\bibitem{yi2020voice}
Z.~Yi, W.-C. Huang, X.~Tian, J.~Yamagishi, R.~K. Das, T.~Kinnunen, Z.-H. Ling,
  and T.~Toda, ``{Voice Conversion Challenge 2020}: Intra-lingual semi-parallel
  and cross-lingual voice conversion,'' in \emph{VCCBC}, 2020.

\bibitem{radford2023robust}
A.~Radford, J.~W. Kim, T.~Xu, G.~Brockman, C.~McLeavey, and I.~Sutskever,
  ``Robust speech recognition via large-scale weak supervision,'' in
  \emph{ICML}, 2023.

\bibitem{das2020predictions}
R.~K. Das, T.~Kinnunen, W.-C. Huang, Z.-H. Ling, J.~Yamagishi, Z.~Yi, X.~Tian,
  and T.~Toda, ``Predictions of subjective ratings and spoofing assessments of
  {V}oice {C}onversion {C}hallenge 2020 submissions,'' in \emph{VCCBC}, 2020.

\bibitem{snyder2018x}
D.~Snyder, D.~Garcia-Romero, G.~Sell, D.~Povey, and S.~Khudanpur, ``X-vectors:
  Robust {DNN} embeddings for speaker recognition,'' in \emph{ICASSP}, 2018.

\bibitem{ITU2015}
{International Telecommunication Union}, ``Method for the subjective assessment
  of intermediate quality level of audio systems,'' 2015.

\bibitem{StatMUSHRA}
C.~Mendon{\c c}a and S.~Delikaris-Manias, ``Statistical tests with {MUSHRA}
  data,'' in \emph{Audio Engineering Society International Convention}, 2018.

\bibitem{Friedman1937}
M.~Friedman, ``The use of ranks to avoid the assumption of normality implicit
  in the analysis of variance,'' \emph{Journal of the American Statistical
  Association}, 1937.

\bibitem{Wilcoxon1945}
F.~Wilcoxon, ``Individual comparisons by ranking methods,'' \emph{Biometrics
  Bulletin}, 1945.

\bibitem{schonemann1966generalized}
P.~H. Sch{\"o}nemann, ``A generalized solution of the orthogonal {Procrustes}
  problem,'' \emph{Psychometrika}, 1966.

\bibitem{ruggiero2025etawavlm}
G.~Ruggiero, M.~Testa, J.~{Van de Walle}, and L.~{Di Caro}, ``{Eta-WavLM}:
  Efficient speaker identity removal in self-supervised speech representations
  using a simple linear equation,'' in \emph{ACL}, 2025.

\end{thebibliography}
    
\end{document}